\documentclass[12pt]{article}
\usepackage{amsmath,amssymb,bm}
\usepackage{graphics}
\usepackage{graphicx}
\usepackage{amssymb}
\usepackage{epstopdf}

\def\pythia{{\sc Pythia}}

\usepackage{mystyle}
\usepackage{cite,./mcite}

\begin{document}

\begin{flushright}
{\small DESY 12-135 } \hspace*{1.9cm}\\
{\small OUTP-12-16P } \hspace*{1.9cm}\\
\end{flushright}

\vskip 1.5 cm 

\begin{center}

{\Large  
Jet production and the inelastic $pp$ cross section \\
at the LHC
}

\vskip 50pt
A.~Grebenyuk$^a$, F. Hautmann$^b$,      
H. Jung$^{a , c}$,  P.~Katsas$^a$ and A.~Knutsson$^a$ \\
\vskip 20pt 
$^a$ Deutsches Elektronen Synchrotron, D-22603 Hamburg\\
$^b$ Theoretical Physics Department, University of Oxford, Oxford OX1 3NP\\
$^c$ Elementaire Deeltjes Fysica, Universiteit Antwerpen, B 2020 Antwerpen\\

\vskip 50pt
\noindent{\bf Abstract}
\vskip 20pt
\end{center}

\noindent{We suggest that,   if  current  measurements of   inclusive  jet production 
 for central rapidities  at the LHC   are     extended 
to   lower   
 transverse momenta,     one could define a   visible  cross section   
  sensitive to  the  unitarity  bound set  by the  recent 
determination  of    the inelastic proton-proton cross section. 
   }

\newpage

Hadronic jet cross sections    are  being explored at the 
Large Hadron Collider (LHC)~\cite{Aad:2011fc,CMS:2011ab}     
across a kinematic range much wider  than at  
any previous collider experiment, with the 
jets  being  measured  either by calorimeters or by trackers. 

The inclusive one-jet cross section at central rapidities  is well described by 
next-to-leading-order QCD calculations and  shower 
 Monte Carlo event generators~\cite{Nason:2012pr}  over   a  wide  range in transverse 
  momenta  from 20 GeV to 2 TeV.  For the case of forward rapidities and 
less inclusive jet observables  see e.g.  discussions in~\cite{Deak:2012rq,Deak:2011ga}.

In this note we suggest that  if  the   jet  measurement  for central 
rapidities  is  extended 
to   lower     but still perturbative   
 transverse momenta  
  one can define a   visible jet cross section   
  sensitive to 
the   bound set  by the inelastic proton-proton rate  which  has recently been 
measured  at the  LHC~\cite{Aad:2011eu,CMS-PAS-QCD-11-002,CMS-PAS-FWD-11-001}.  
We observe that this can be done  within the range of 
acceptance of the measurement without using any extrapolation. 

We start by briefly recalling the physical picture 
and Monte Carlo estimates~\cite{Sjostrand:2006za,Sjostrand:2004pf,Gustafson:2002kz,Gustafson:2002jy,Gustafson:1999kh} 
for the  increase   in  the partonic cross section  
at low transverse momenta.  Then we  consider  measuring   jets  
in the  low-momentum  region,  and 
introduce the visible jet cross section. We present numerical estimates and comment 
on physical implications of the proposed measurements.

To begin,   consider  the basic  physical picture 
 for  jet production in the high-energy limit.  In this limit 
  the dynamics is driven by   the growth of gluon densities at low 
  momentum fractions $ x \sim  (  p_T /  \sqrt{s} )   e^{-y} $, where 
  $p_T$ and $y$ are the jet transverse momentum and rapidity,   and $\sqrt{s}$ is the 
  center-of-mass energy. 
  As the energy increases, the jet cross section rises, and eventually 
 the perturbative prediction obtained from integrating  the cross section over  
  transverse momenta  above a given   $p_T$ 
 is higher than the inelastic pp cross section.  
 The value of  $p_T$  at which this occurs depends on the 
 parton density and the center-of-mass energy.  At the LHC,  for the first time,    
 such a  $p_T$  value  approaches  the weakly coupled region,  
$p_T  = {\cal O } (10)$ GeV~\cite{Sjostrand:2004pf},  owing to the high center-of-mass energy and 
 the associated  wide rapidity phase space.  

Dynamical effects slowing down the rise of the 
  unitarity-violating  cross section 
   go beyond the  QCD parton model approximations  
  valid at large transferred   momenta. Even though at weak coupling, 
  they   involve strong fields and non-perturbative  physics. 
 At phenomenological level, 
 this    constitutes    the motivation for 
 multi-parton interaction models~\cite{Sjostrand:2006za,Sjostrand:2004pf,Gustafson:2002kz,Gustafson:2002jy,Gustafson:1999kh}  in 
 shower Monte Carlo event generators for  event  simulation of  final states.  
 However,  fundamental  aspects  of these processes are  little  known. 
  In addition, although all collider  Monte Carlo models  appear to require 
  multi-parton interactions to explain  various  features of  pp data, the experimental evidence for double parton scatterings is still weak.
  
We here  propose  appealing  to   
the measurements~\cite{Aad:2011eu,CMS-PAS-QCD-11-002,CMS-PAS-FWD-11-001}  
 to  investigate    the   jet cross section near the $p_T$  region where 
 the inelastic pp production rate is saturated. 
This region  requires  jets at low $p_T$ and therefore only jets constructed from charged tracks can experimentally be employed.
In addition, forward  high-$p_T$ 
 particle  production is copious at the LHC and poses special issues.   
 On the experimental side, 
particle tracking capabilities  decay   with increasing  rapidity~\cite{Jung:2009eq}. 
On the theory side,  QCD  predictions are affected by 
  all-order  logarithmic corrections~\cite{Deak:2012rq,Deak:2011ga,Deak:2009xt,Deak:2010gk} increasing with rapidity. 
We thus focus on the central  pseudorapidity  range. In this region the {\it visible} inelastic cross section has been measured by ATLAS and CMS \cite{Aad:2011eu,CMS-PAS-QCD-11-002,CMS-PAS-FWD-11-001} giving a value of $\sigma_{inel} \sim 60$~mb at $\sqrt{s}=7$~TeV, depending on the precise definition of the visible phase space. No extrapolation is needed for this to be compared with the jet cross section proposed here.

\begin{figure}[htbp]
\includegraphics[scale=0.4]{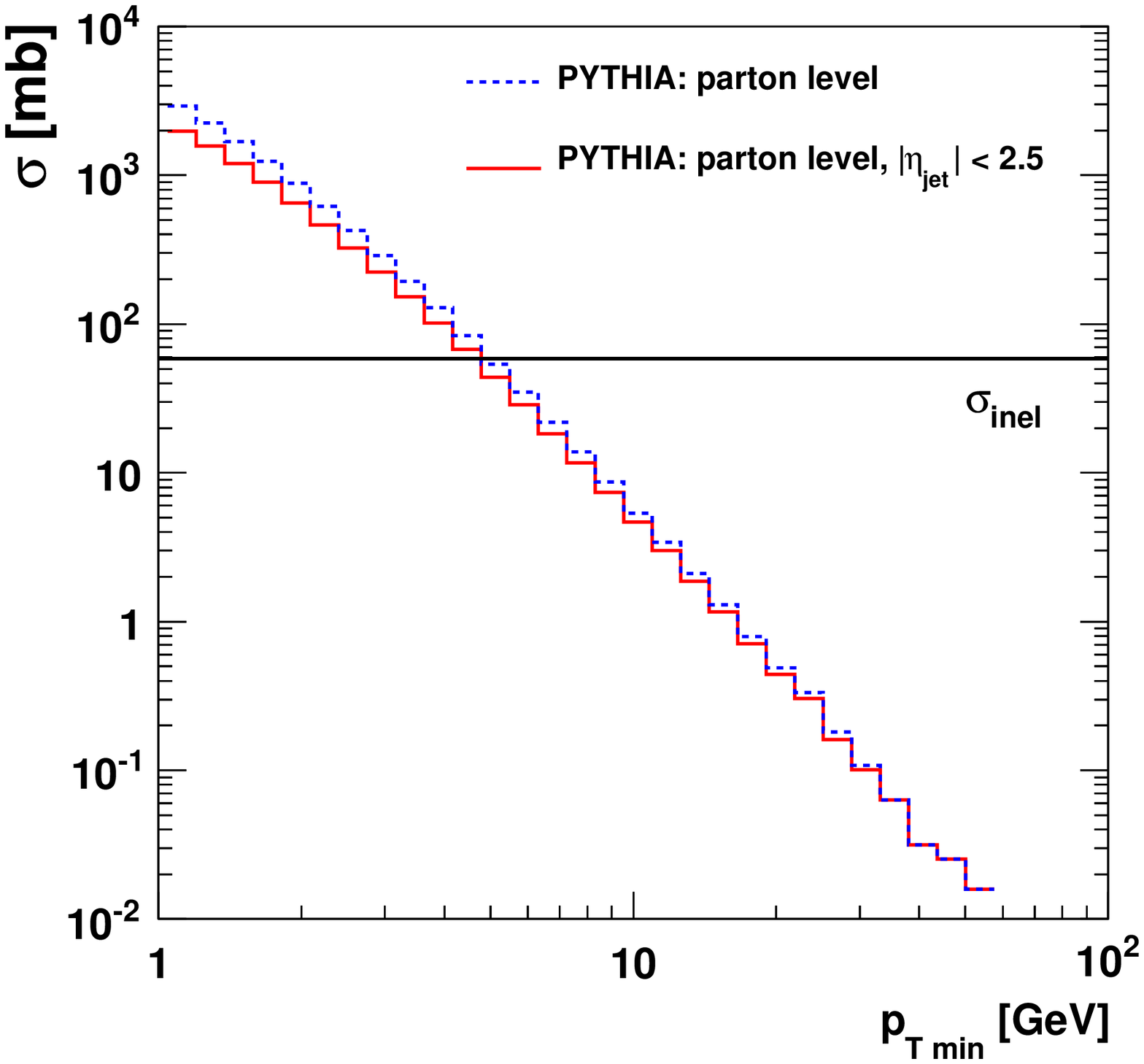}
\includegraphics[scale=0.4]{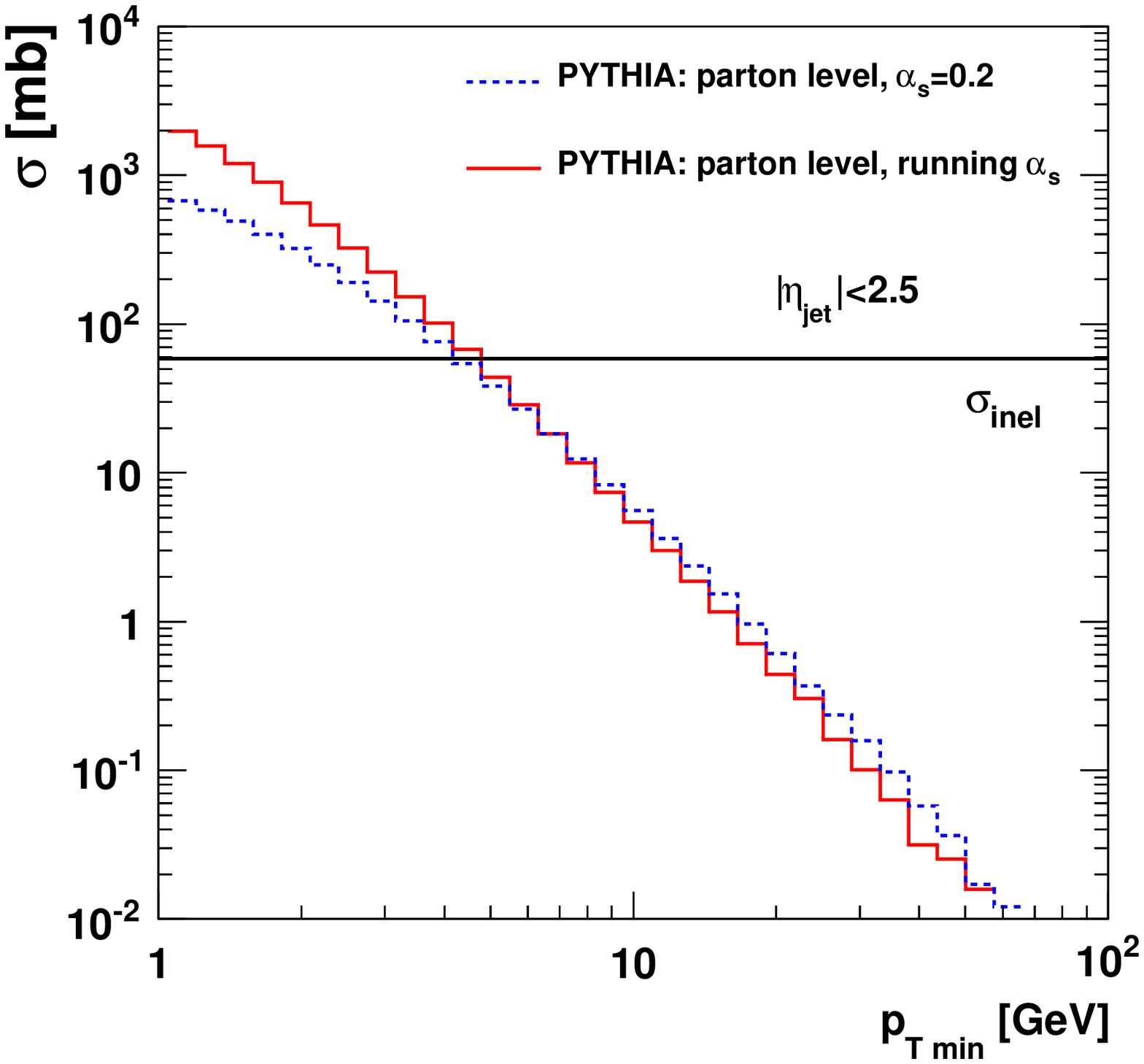}
\caption{\it  Integrated  cross sections at $\sqrt{s}=8$~TeV as a function of 
the minimum transverse momentum:  (left) inclusive cross section compared to what can be investigated within $|\eta|<2.5$;
(right) visible inclusive cross section in $|\eta|<2.5 $ compared to the prediction with fixed $\alpha_s = 0.2$.}  
\label{fig:fig1}
\end{figure}

First we consider the  parton-level  cross section at $\sqrt{s}=8$~TeV (calculated using \pythia\   (version 6.425) \cite{Sjostrand:2006za}).  Figure~\ref{fig:fig1} (left) 
shows  the estimate obtained 
from the $ 2 \to 2$  integrated cross section 
 as a function of the minimum transverse momentum:
 \begin{equation}
\sigma(p_{T\;min}) = \int_{p_{T\;min}}dp^2_T  \int_{-\infty}^{\infty} dy  \frac{d^2 \sigma}{dp^2_T\,dy}  = \int_{p_{T\;min}}dp^2_{T\,jet}  \int_{-\infty}^{\infty} dy_{jet}  \frac{d^2 \sigma_{jet}}{dp^2_{T\,jet} \,dy_{jet}} \, ,
\label{eq2}
\end{equation}
where the last expression gives an operational definition of $\sigma(p_{T\;min}) $ in terms of a measureable  leading jet cross section $\sigma_{jet}$.
In   Fig.~\ref{fig:fig1} (left)   we define  
the visible range  by restricting 
 the integration to  the  pseudorapidity region 
$ | \eta | <  2.5 $.
For comparison we plot the measurement~\cite{Aad:2011eu,CMS-PAS-QCD-11-002,CMS-PAS-FWD-11-001} of the inelastic 
cross section as a horizontal line. 
One can clearly see that the partonic cross section exceeds the inelastic cross section at values of the transverse momentum at around 4 -- 5 GeV even in the restricted $\eta$ range. In Fig.~\ref{fig:fig1}  (right) we also show the cross section with $|\eta|<2.5$ using a fixed value of $\alpha_s=0.2$. This
illustrates  that the infrared behaviour of the strong  coupling  does not affect significantly the  physical picture  in the $p_T$  region where the  jet cross section   approaches the inelastic bound.  
The rise of the cross section is essentially coming from the $1/t^2$ pole of the partonic matrix element, as explained in \cite{Sjostrand:2004pf}.

\begin{figure}[htbp]
\includegraphics[scale=0.4]{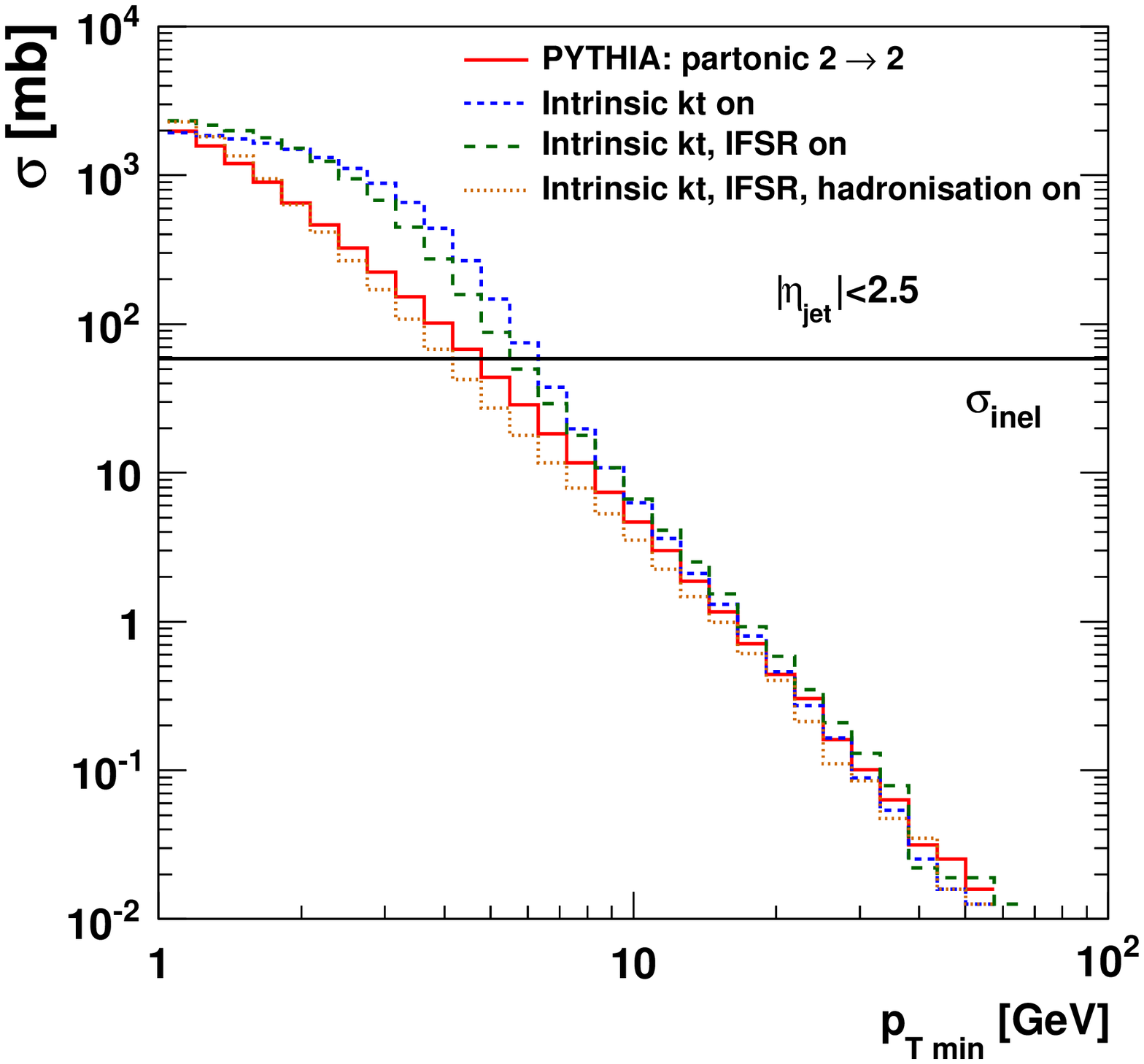}
\includegraphics[scale=0.4]{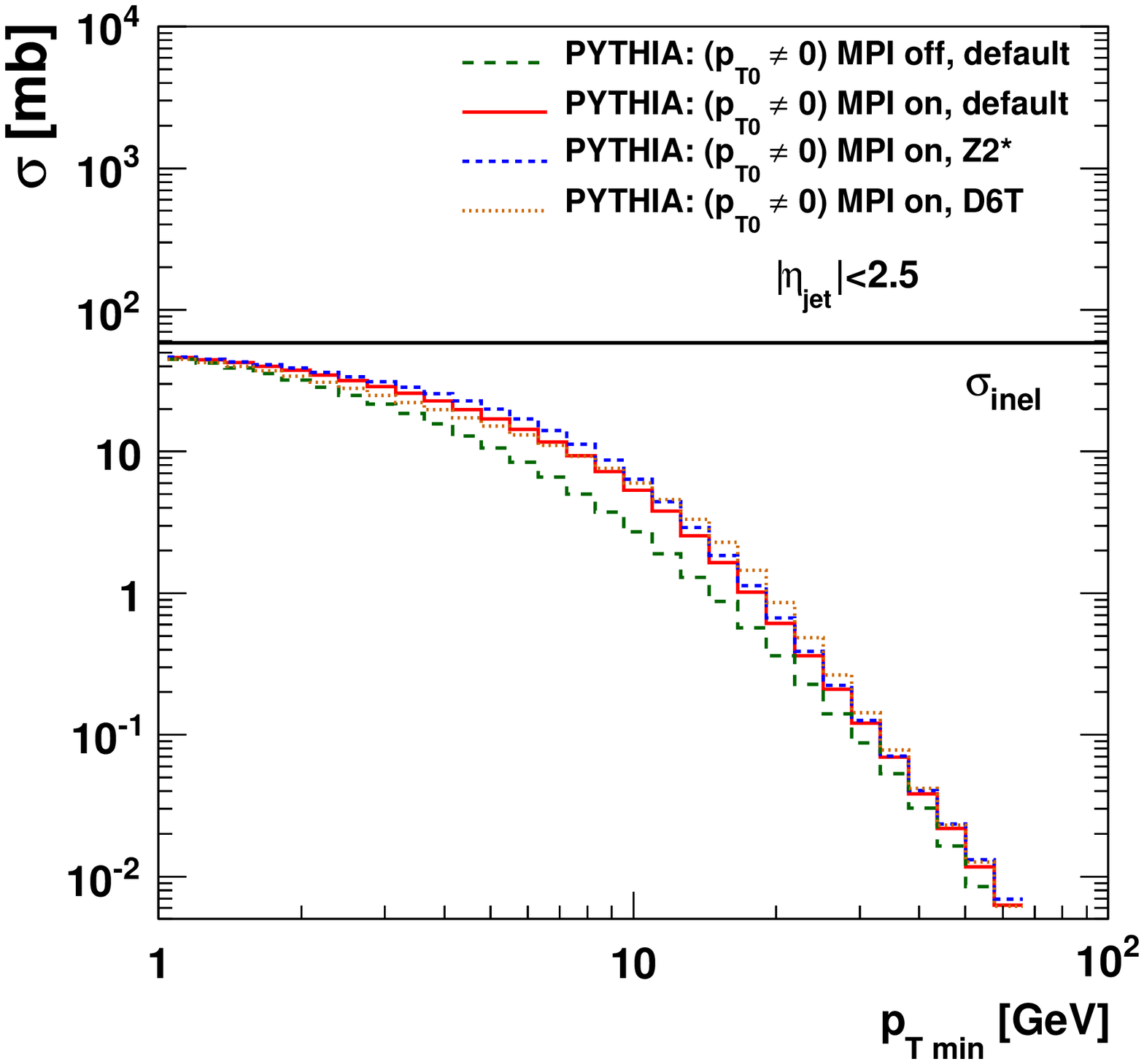}
\caption{\it (Left) Cross section from purely partonic $2\to2$ process, including intrinsic $k_t$-effects, including initial and final state parton showers (IFSR) and finally hadronisation.
(Right) predicted cross section applying $p_{T0}\neq 0 $ and MPI with different underlying event tunes of \protect\pythia . }  
\label{fig:fig2}
\end{figure}

We  then  consider the cross section of jets at particle level.   We suppose   measuring  jets  
at  low transverse momenta  in the visible range   $ | \eta | <  2.5 $. 
In order to  reconstruct the jets     we    use   the  
anti-k$_T$    algorithm~\cite{Cacciari:2008gp} with $R=0.5$ down to    low transverse momenta.
The visible jet cross section is shown in Fig.~\ref{fig:fig2} using \pythia~\cite{Sjostrand:2006za}. The solid line  corresponds to the partonic cross section (of Fig.~\ref{fig:fig1}). We show the effect of turning on successively intrinsic $k_t$, initial and final state parton showers (IFPS) and finally hadronisation (using default parameters, without allowing a taming of the cross section). 
In Fig.~\ref{fig:fig2} (left)   the  perturbative result  reaches the 
inelastic bound~\cite{Aad:2011eu} for minimum $ p_T \simeq 4 $ GeV. 
In  the region  just above this value, $p_T  = {\cal O } (10)$ GeV,   effects  responsible for the taming of the cross section set in.  
The model~\cite{Sjostrand:1987su,Sjostrand:2004pf} provides a phenomenological modification of the low-$p_T$  behaviour of the jet  cross section within a collinearly-factorised framework; the rise of the cross section is tamed at small values of $p_T$ by introducing a factor:
\begin{equation}
\frac{\alpha^2_s(p^2_{T0} + p_T^2)}{\alpha^2_s(p^2_{T})}\frac{p^4_T}{(p^2_{T0}+p_T^2)^2} \; ,
\label{ref:eq}
\end{equation}
where $p_{T0}$ is a parameter obtained from  a fit to describe measurements of the underlying event. 
 In Fig.~\ref{fig:fig2} (right) we show the cross section based on eq.(\ref{ref:eq}) as well as the effect of multi-parton interactions (MPI). The prediction of tune D6T~\cite{D6T} and Z2* are also shown. The Z2* tune is an updated version of the Z1 tune~\cite{Field:2010bc}, based on automated tuning and using the CTEQ6L PDF~\cite{Pumplin:2002vw}.  In the Z2* tune PARP(82)=1.92 and PARP(90)=0.23.

Note that in approaches that 
go beyond the collinear approximation~\cite{Gustafson:2002kz,Gustafson:2002jy,Gustafson:1999kh,Deak:2009xt,Deak:2010gk,Hautmann:2008vd,Hautmann:2007gw,Hautmann:2009zzb}  
the low-$p_T$  behaviour results from  
two different sources: first, the perturbative matrix elements, which are computed 
at finite transverse momenta $k_T$ in the  initial state,   have the standard collinear 
rise~\cite{Sjostrand:2006za} 
 at low $p_T$ for $k_T  \ll   p_T$ but a slower rise for 
 $k_T  \simeq   p_T$~\cite{Deak:2009xt,Deak:2010gk,Hautmann:2007gw,Hautmann:2009zzb};  second, the unintegrated parton  
 densities  enhance  the relative weight of  finite transverse momentum contributions 
 compared to collinearly-ordered contributions,  due  to    both Sudakov and Regge 
  suppression of the     low-$k_T$ region~\cite{Gustafson:2002kz,Gustafson:2002jy,Hautmann:2007gw,Hautmann:2009zzb}.  
The jet measurements suggested in this paper may be useful to 
 investigate    $k_T$-dependent dynamical effects. 
\begin{figure}[htbp]
\includegraphics[scale=0.4]{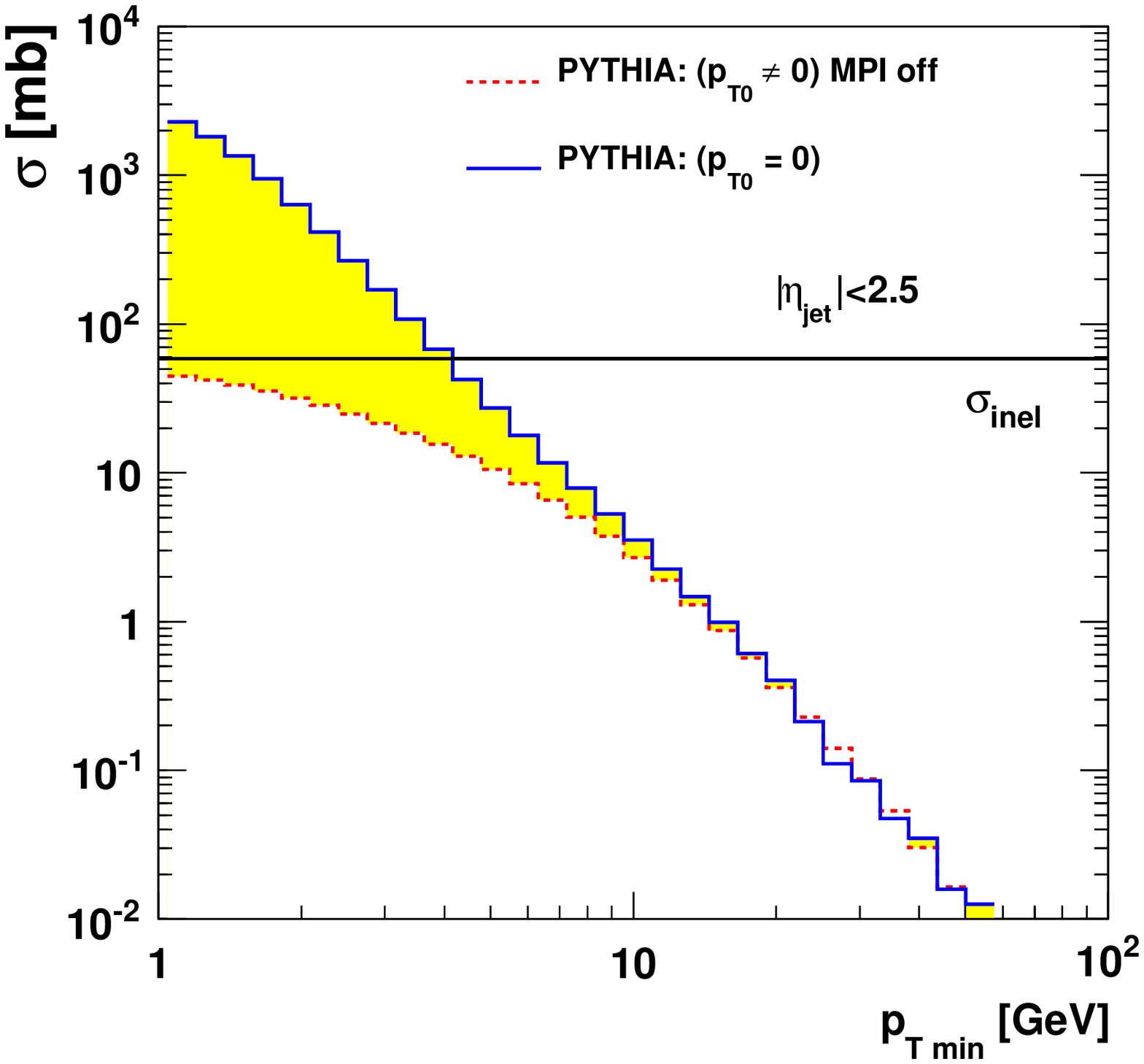}
\includegraphics[scale=0.4]{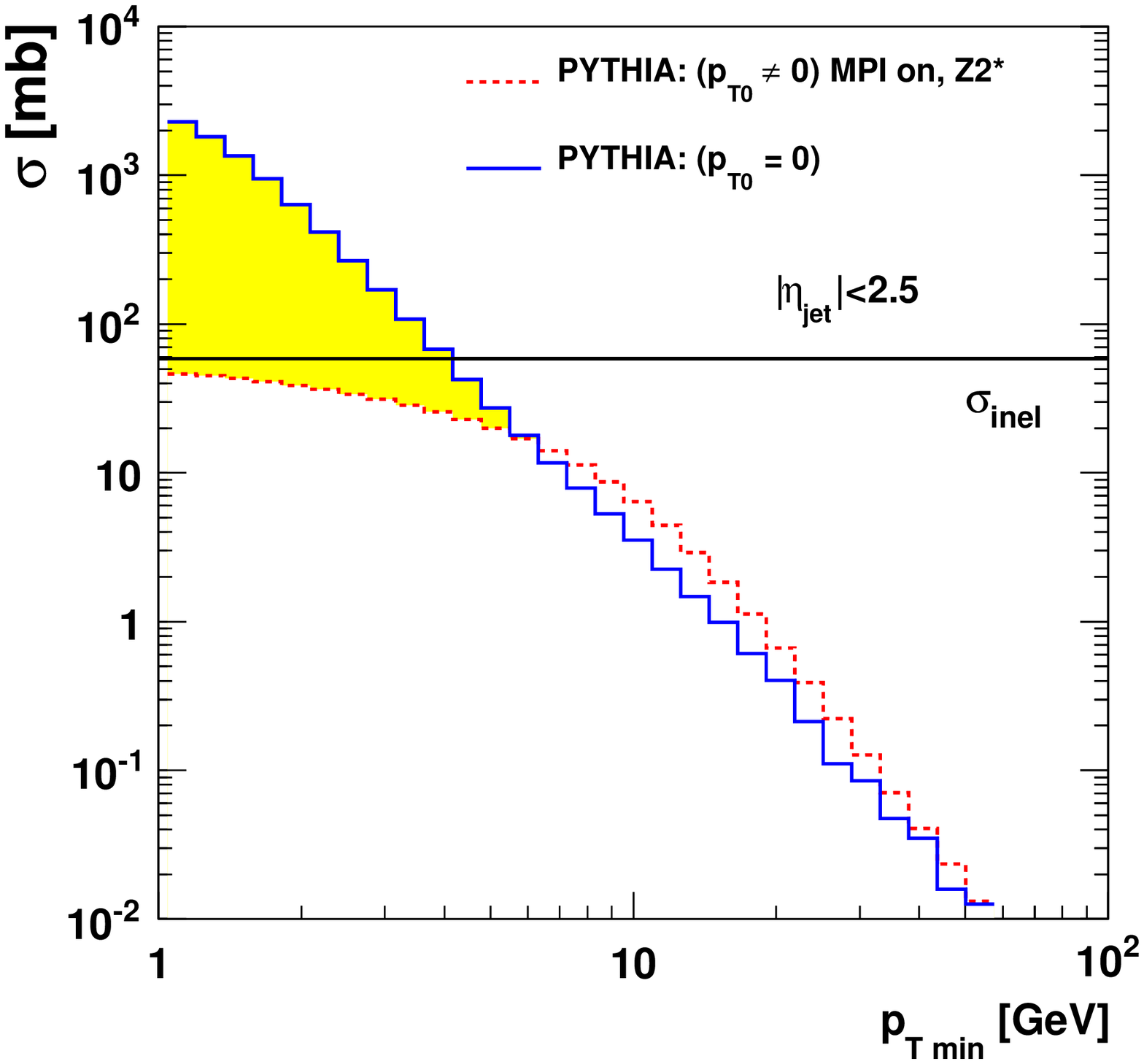}
\caption{\it The cross section as a function of $p_{T\,min}$ as predicted by  \protect\pythia ~in the range $|\eta| < 2.5$. The solid (blue) line shows the prediction applying
$p_{T0}=0$ including parton shower and hadronisation, while the dashed (red) line shows the prediction with $p_{T0}\neq 0$;
(left) is without multi-parton interactions, (right)  including multi-parton interactions with tune Z2* ~\protect\cite{Field:2010bc}.}
\label{fig:fig3}
\end{figure} 
In Fig.~\ref{fig:fig3} we show a comparison of the jet cross section for $p_{T0}=0$, including parton shower and hadronisation,  with the cross section obtained from \pythia    
 ~including the model of eq.(\ref{ref:eq}). In Fig.~\ref{fig:fig3} (right) we show the effect  of multi-parton interactions. Especially in the region of $p_T < 10$~GeV a clear deviation from the $p_{T0}=0$ prediction is visible. A measurement of the jet cross section in the low $p_T$ region would give insight into the transition from the perturbative jet cross section at large $p_T$ to the small $p_T$ region where additional effects are needed to avoid unitarity violation.
Besides $pp$ collisions, the jet measurements proposed
may be extended to collisions of nuclei. If the inelastic
cross section is measured in AA and pA, they may be useful
to characterise properties of final states in terms of jets or
flows, and investigate the role of $k_T$-dependent effects
and multiple interactions in ion collisions.

In summary, we have proposed measurements of the visible jet cross section 
at the LHC by using jets  down to  transverse momenta  of the order of a few GeV. 
Such  measurements require a handle on jet algorithms and jet 
reconstruction capabilities in the low $p_{T}$ region. 
Without  extrapolation, they could be 
related to the measurement~\cite{Aad:2011eu,CMS-PAS-QCD-11-002,CMS-PAS-FWD-11-001}    of the  inelastic $pp$ cross section. 
The jet cross section measurement in the proposed  $p_T$ region 
probes   weak-coupling but still non-perturbative   QCD physics.  It is 
relevant for phenomenology of  multi-parton interaction models in  shower Monte Carlo 
generators.  It can  provide new experimental information on 
transverse momentum dependent 
correlations  between  initial and final states of 
partonic collisions.   

\section*{Acknowledgements}
We are grateful for many instructive discussions with T. Sj\"ostrand and P. Skands.


\begin{thebibliography}{10}

\bibitem{Aad:2011fc}
ATLAS, G.~Aad {\em et~al.},
\newblock Phys. Rev. {\bf D86}, 014022 (2012), arXiv:1112.6297.

\bibitem{CMS:2011ab}
CMS, S.~Chatrchyan {\em et~al.},
\newblock Phys.Rev.Lett. {\bf 107}, 132001 (2011), arXiv:1106.0208.

\bibitem{Nason:2012pr}
P.~Nason and B.~Webber,
\newblock (2012), arXiv:1202.1251.

\bibitem{Deak:2012rq}
M.~Deak, F.~Hautmann, H.~Jung, and K.~Kutak,
\newblock (2012), arXiv:1206.7090.

\bibitem{Deak:2011ga}
M.~Deak, F.~Hautmann, H.~Jung, and K.~Kutak,
\newblock Eur.Phys.J. {\bf C72}, 1982 (2012), arXiv:1112.6354.

\bibitem{Aad:2011eu}
ATLAS, G.~Aad {\em et~al.},
\newblock Nature Commun. {\bf 2}, 463 (2011), arXiv:1104.0326.

\bibitem{CMS-PAS-QCD-11-002}
\mbox{CMS Collaboration},
\newblock {Measurement of the inelastic pp cross section at sqrts = 7 TeV},
\newblock {CMS-PAS-QCD-11-002 https://cdsweb.cern.ch/record/1433413?ln=en},
  2012.

\bibitem{CMS-PAS-FWD-11-001}
\mbox{CMS Collaboration},
\newblock {Inelastic pp cross section at 7 TeV},
\newblock {CMS-PAS-FWD-11-001 https://cdsweb.cern.ch/record/1373466?ln=en},
  2011.

\bibitem{Sjostrand:2006za}
T.~Sj\"ostrand, S.~Mrenna, and P.~Skands,
\newblock JHEP {\bf 05}, 026 (2006), arXiv:hep-ph/0603175.

\bibitem{Sjostrand:2004pf}
T.~Sj\"ostrand and P.~Skands,
\newblock JHEP {\bf 03}, 053 (2004), arXiv:hep-ph/0402078.

\bibitem{Gustafson:2002kz}
G.~Gustafson, L.~L\"onnblad, and G.~Miu,
\newblock Phys.Rev. {\bf D67}, 034020 (2003), arXiv:hep-ph/0209186.

\bibitem{Gustafson:2002jy}
G.~Gustafson, L.~L\"onnblad, and G.~Miu,
\newblock JHEP {\bf 0209}, 005 (2002), arXiv:hep-ph/0206195.

\bibitem{Gustafson:1999kh}
G.~Gustafson and G.~Miu,
\newblock Phys.Rev. {\bf D63}, 034004 (2001), arXiv:hep-ph/0002278.

\bibitem{Jung:2009eq}
Z.~J. Ajaltouni {\em et~al.},
\newblock (2009), arXiv:0903.3861.

\bibitem{Deak:2009xt}
M.~Deak, F.~Hautmann, H.~Jung, and K.~Kutak,
\newblock JHEP {\bf 09}, 121 (2009), arXiv:0908.0538.

\bibitem{Deak:2010gk}
M.~Deak, F.~Hautmann, H.~Jung, and K.~Kutak,
\newblock {Forward-Central Jet Correlations at the Large Hadron Collider},
  2010, arXiv:1012.6037.

\bibitem{Cacciari:2008gp}
M.~Cacciari, G.~P. Salam, and G.~Soyez,
\newblock JHEP {\bf 04}, 063 (2008), arXiv:0802.1189.

\bibitem{Sjostrand:1987su}
T.~Sj\"ostrand and M.~van Zijl,
\newblock Phys. Rev. {\bf D36}, 2019 (1987).

\bibitem{D6T}
R.~Field,
\newblock Studying the underlying event at {CDF} and the {LHC},
\newblock in {\em {Multiple partonic interactions at the LHC. Proceedings, 1st
  International Workshop, MPI'08, Perugia, Italy, October 27-31, 2008}}, edited
  by P.~Bartalini and L.~Fan{\'o}, pp. 12--31, 2010, arXiv:1003.4220.

\bibitem{Field:2010bc}
R.~Field,
\newblock (2010), arXiv:1010.3558.

\bibitem{Pumplin:2002vw}
J.~Pumplin {\em et~al.},
\newblock JHEP {\bf 0207}, 012 (2002), arXiv:hep-ph/0201195.

\bibitem{Hautmann:2008vd}
F.~Hautmann and H.~Jung,
\newblock JHEP {\bf 10}, 113 (2008), arXiv:0805.1049.

\bibitem{Hautmann:2007gw}
F.~Hautmann and H.~Jung,
\newblock Nucl.Phys.Proc.Suppl. {\bf 184}, 64 (2008), arXiv:0712.0568.

\bibitem{Hautmann:2009zzb}
F.~Hautmann,
\newblock Acta Phys.Polon. {\bf B40}, 2139 (2009).

\end{thebibliography}
\end{document}